\documentclass{PoS}

\title{\begin{picture}(0,0)(0,0)
  \put(420,110){\makebox(0,0)[l]{\textnormal{\normalsize
 \textbf{ CHIBA-EP-235}}}}
\end{picture}Composite operator and condensate in $SU(N)$ Yang-Mills theory with $U(N-1)$ stability group}

\ShortTitle{Condensate in a novel decomposition of Yang-Mills theory}

\author{\speaker{Matthias Warschinke}\thanks{Based on \cite{Warschinke:2017fgw} %in collaboration with Ryutaro Matsudo, Shogo Nishino, Toru Shinohara, Kei-Ichi Kondo.
}\\
        Department of Physics, Graduate School of Science, Chiba University\\
        E-mail: \email{m\_warschinke@chiba-u.jp}}

\author{Ryutaro Matsudo\\
        Department of Physics, Graduate School of Science and Engineering,
Chiba University\\
        E-mail: \email{afca3071@chiba-u.jp}}
        
        \author{Shogo Nishino\\
       Department of Physics, Graduate School of Science, Chiba University\\
        E-mail: \email{shogo.nishino@chiba-u.jp}}
        
        \author{Toru Shinohara\\
        Department of Physics, Graduate School of Science, Chiba University\\
        E-mail: \email{sinohara@graduate.chiba-u.jp}}
        
        \author{Kei-Ichi Kondo\\
        Department of Physics, Graduate School of Science / and Engineering,
Chiba University\\
        E-mail: \email{kondok@faculty.chiba-u.jp}}

\abstract{Recently, a reformulation of the $SU(N)$ Yang-Mills theory inspired by the
Cho-Faddeev-Niemi decomposition has been developed in order to
understand confinement from the viewpoint of the dual superconductivity. The concept of infrared Abelian dominance plays an important role in the realization of this concept and through numerical simulations on the lattice, evidence was found for example in the form of the dynamical mass generation for certain gluon degrees of freedom. A promising analytical attempt to explain the generation of such masses is through condensates of mass dimension two. 
In this talk, we want to focus on the reformulated $SU(N)$ Yang-Mills theory in
the previously overlooked minimal option with the non-Abelian $U(N-1)$ stability group, in contrast to the famous maximal Abelian gauge, where the decomposition corresponds to the Abelian $U(1)^{N-1}$ stability group. We proceed with a thorough one-loop analysis of this novel decomposition, calculating all standard renormalization group functions at one-loop level in light of the renormalizability of this theory. We subsequently define an appropriate  mixed gluon-ghost composite operator of mass dimension two as the candidate for the condensate within this theory and prove its (on-shell) BRST invariance and the multiplicative renormalizability. Finally, the existence of the condensate is discussed within the local composite operator formalism.}

\FullConference{XIII Quark Confinement and the Hadron Spectrum - Confinement2018\\
		31 July - 6 August 2018\\
		Maynooth University, Ireland}

\usepackage{amsmath,amsfonts,amssymb,amsthm,mathrsfs}

\newcommand{\muu}[1]{\begin{align} #1 \end{align}}

\newcommand{\mmo}[1]{\begin{pmatrix} #1 \end{pmatrix}}

\DeclareMathOperator{\Tr}{Tr}
\newcommand{\zogh}{Z_{\omega}}

\newcommand{\zdia}{Z_C}
\newcommand{\zadia}{Z_{\bar{C}}}

\usepackage{tikz}
\usetikzlibrary{shapes.geometric, arrows}

\tikzstyle{startstop} = [rectangle, minimum width=3cm, minimum height=0.7cm, text centered, text width=3cm, draw=black, fill=orange!30]

\tikzstyle{arrow} = [thick,->,>=stealth]

\definecolor{english}{rgb}{0.0, 0.5, 0.0}

\begin{document}

\section{Dual superconductivity and reformulation of Yang-Mills theory}

Understanding the phenomenon of (quark-) confinement remains one of the biggest challenges within theoretical physics, despite the existence of several promising approaches. One of them is the dual superconductivity picture due to 't Hooft, Nambu and Mandelstam \cite{Nambu:1974zg,thooft:bologna,Mandelstam:1974pi,tHooft:1981bkw}. The dual superconductivity picture advocates the idea of a superconducting Yang-Mills vacuum consisting of a color magnetic monopole condensate. If we put any color-electrically charged objects into such a vacuum, for example a static quark-antiquark pair, the equivalent of the dual Meissner effect would confine the color electric field into nearly one-dimensional objects, the flux tubes, leading to a linearly rising potential between the quarks. On the other hand, however, the dual Landau-Ginzburg model describing such a dual superconductor is an Abelian theory. A procedure to extract the monopole degrees of freedom is given by the ``Abelian projection'' \cite{tHooft:1981bkw}, a procedure of gauge-fixing the non-Abelian degrees of freedom. A famous representative is the maximal Abelian gauge (MAG), where the gauge fixing condition relates to minimizing those degrees of freedom that correspond to the off-diagonal generators of the gauge group $SU(N)$. Within this setting, lattice simulations showed that these degrees of freedom become massive in the infrared \cite{Amemiya:1998jz}, which can be seen as a confirmation of the hypothesis of ``infrared Abelian dominance'' \cite{Ezawa:1982bf}. Still, two problems remain.

First, the fact that the identification of the monopoles relies on some sort of gauge fixing lead to the criticism of monopoles being mere gauge artifacts. This situation was improved through a certain reformulation of Yang-Mills theory \cite{Kondo:2005eq,Kondo:2005ge,Kondo:2008xa} based on the Cho-Duan-Ge-Faddeev-Niemi-Shabanov decomposition of the gauge field with respect to the stability group $H$ of the gauge group $G=SU(N)$,
\begin{equation}
\mathcal{A}_\mu = \mathcal{X}_\mu + \mathcal{V}_\mu \in \ \text{Lie}\ G/H \oplus \text{Lie}\ H.
\end{equation}
The decomposition is defined in terms of the color-field $\mathfrak{n}(x)$. However, from a new standpoint, this decomposition is regarded a non-linear change of variables $\{\mathcal{A}_\mu\}\to \{\mathcal{X}_\mu,\mathcal{V}_\mu,\mathfrak{n}\}$. The theory in the new variables not only possesses an enlarged gauge symmetry by rotations along the color-field axis $SU(N)\to [SU(N)]_\mathcal{A}\times [SU(N)/H]_\mathfrak{n}$, but also exceeds the original degrees of freedom. By imposing the so called reduction condition,
\begin{equation}
\mathcal{D}_\mu[\mathcal{V}]\mathcal{X}^\mu = 0,
\end{equation}
the enlarged symmetry is reduced to the original $SU(N)$ symmetry and the superfluous degrees of freedom are removed.

The reduction condition is incorporated into the path-integral via the Faddeev-Popov trick, but is conceptually different from the usual gauge fixing, as the local $SU(N)$ gauge symmetry is still intact and the usual gauge fixing procedure must still be performed. In this prescription, the infrared dominant degrees of freedom encoded in the residual (or restricted) field $\mathcal{V}_\mu$ and the coset (or remaining) field $\mathcal{X}_\mu$, which is expected to become massive, can be extracted in a gauge invariant manner. Another important result of these investigations was that the decomposition is not unique, but allows different choices for the stability group, at least for $SU(N)$ with $N>2$. The from the viewpoint of QCD interesting case of $SU(3)$ enables two choices, the maximal option with $H^{max}=U(1)\times U(1)$, which is related to the MAG, and a previously overlooked choice called minimal option, $H^{min}=U(2)$. It is the latter one on which we are going to focus in this talk.

The second remaining problem is to find an analytical explanation for the generation of the masses observed in the lattice simulations, i.e., an analytical description of the infrared Abelian dominance. This is where the condensates of mass dimension two come into play. In fact, an effective mass term could for example be obtained from the quartic gluon interaction, if such condensates exist. A promising candidate among the condensates was first proposed in \cite{Kondo:2001nq} and is related to the mixed gluon-ghost composite operator,
\muu{
\mathcal{O}=\Tr_{G/H}\left(\mathcal{X}_\mu \mathcal{X}^\mu -2 i \xi \mathcal{C}\bar{\mathcal{C}} \right), \label{operator}
} 
where $\xi$ is the ``gauge fixing'' parameter related to the reduction condition. In this talk, we wish to examine the condensation of this operator rather than the gauge-invariant extraction of the monopoles. Therefore, we perform an approximation by fixing the color-field to be identical to the last Cartan generator, $\mathfrak{n}\equiv T^{N^2-1}$. This breaks the symmetry in the sense that the reduction condition no longer preserves full $SU(N)$ gauge symmetry, but leads to a breaking $SU(N) \to H=U(N-1)$\footnote{In the same way, the MAG can be recovered from the reformulated theory if one performs the decomposition with $H^{max}$.}. The residual gauge symmetry is fixed in the standard Lorenz gauge, however, from viewpoint of renormalizability it turned out that we further need to decompose $U(N-1)=SU(N-1)\times U(1)$. The theory under consideration therefore becomes,
\muu{
\mathcal{L}=\mathcal{L}_{YM}+\mathcal{L}_{GF+FP}^{RED}+\mathcal{L}_{GF+FP}^{RES} \label{theory},
}
with,
\muu{
\mathcal{L}_{GF+FP}^{RED}&= i\delta_B \bar{\delta}_B \Tr_{G/H}\left(\mathcal{X}_\mu \mathcal{X}^\mu - i \xi \mathcal{C}\bar{\mathcal{C}} \right), \nonumber \\
\mathcal{L}_{GF+FP}^{RES}&=-i\delta_B  \Tr_{U(1)}\left(\bar{\mathcal{C}}\left[\partial_\mu \mathcal{V}^\mu +\frac{\alpha}{2}\mathcal{N}\right] \right)-i\delta_B  \Tr_{SU(N-1)}\left(\bar{\mathcal{C}}\left[\partial_\mu \mathcal{V}^\mu +\frac{\lambda}{2}\mathcal{N}\right] \right).
}
Here, $\mathcal{N}$ denotes the Nakanishi-Lautrup field and $\delta_B$ and $\bar{\delta}_B$ are the BRST and anti-BRST transformations, respectively. 

\section{Renormalizability of the theory}

As a first step, we perform the one-loop renormalization of the theory \eqref{theory} and investigate the multiplicative renormalizability of the composite operator. The Yang-Mills part of the Lagrangian decomposes according to,
\begin{align}
\mathcal{L}_{YM}=-\frac{1}{4} F_{\mu \nu}^a F^{\mu \nu a} - \frac{1}{4}F_{\mu \nu}^j F^{\mu \nu j}- \frac{1}{4}F_{\mu \nu}^\gamma F^{\mu \nu \gamma},
\end{align}
where the field strength tensors of the various sectors are given by,
\begin{align}
F_{\mu \nu}^a = &D_\mu^{ab} X_\nu^b - D^{ab}_{\nu} X_\mu^b;\ \ \ \ \ \ \ \ F_{\mu \nu}^\gamma = \partial_\mu V_\nu^\gamma - \partial_\nu V_\mu^\gamma +g f^{\gamma ab} X_\mu^a X_\nu^b,\nonumber \\
&F_{\mu \nu}^j = \partial_\mu V_\nu^j - \partial_\nu V_\mu^j +g f^{jab} X_\mu^a X_\nu^b+gf^{jkl}V_\mu^k V_\nu^l,
\end{align}
with the covariant derivative $D_\mu^{ab} \equiv D_\mu^{ab}[V] = \delta^{ab} \partial_\mu + gf^{aJb}V_\mu^J$. We labelled the $2(N-1)$ coset degrees of freedom by $a,b,...$, the $N(N-2)$ elements of the $SU(N-1)$ sector by $j,k,...$ and finally $\gamma \equiv N^2-1$ labels the $U(1)$ sector. For convenience, the latter ones are sometimes combined into $J,K,...$ labelling the full $U(N-1)$ sector. The corresponding generators satisfy the basic commutator relations,
\muu{
[T^a,T^b]=if^{abJ}T^J; \ \ [T^a,T^J]=if^{aJc}T^c; \ \ [T^j,T^k]=if^{jkl}T^l; \ \ [T^j,T^\gamma]=0.
}
After integrating out the Nakanishi-Lautrup field, the gauge fixing Lagrangian is cast into the form,
\begin{align}
\mathcal{L}_{GF+FP}^{RED}&=-\frac{1}{2\xi}(D_\mu^{ab}X^{\mu b})^2 +i\bar{\omega}^a D^{\mu ab}D_\mu^{bc}\omega^c \nonumber \\
&+\frac{\xi g^2}{4} f^{abJ}f^{cdJ} \bar{\omega}^a \bar{\omega}^b \omega^c \omega^d +ig^2 f^{abJ}f^{cdJ}  X^{\mu a} X_\mu^c \bar{\omega}^b \omega^d. \label{LRED}\\
\mathcal{L}_{GF+FP}^{RES}&=-\frac{1}{2\lambda}\left( \partial_\mu V^{\mu j}\right)^2-\frac{1}{2\alpha}\left( \partial_\mu V^{\mu \gamma}\right)^2+i \bar{C}^J \partial^2 C^J\nonumber \\
&  +igf^{jkl}\bar{C}^j \partial_\mu (V^{\mu k}C^l)+igf^{Jab}\bar{C}^J \partial_\mu (X^{\mu a}\omega^b),\label{LRES} 
\end{align}
Here, $\omega$ and $\bar{\omega}$ denote the coset ghosts while $C$ and $\bar{C}$ denote the residual ghosts. Upon introducing the following set of renormalization constants,

\begin{minipage}{.19\textwidth}
\begin{align*}
X_\mu^a &= Z_X^{\frac{1}{2}} X_{\mu R}^a,\\
C^j &= \zdia^\frac{1}{2} C^j_R,\\
\phantom{x}&\phantom{=}
\end{align*}
\end{minipage}
\begin{minipage}{.19\textwidth}
\begin{align*}
V_\mu^j &= Z_V^{\frac{1}{2}} V_{\mu R}^j,\\
\bar{C}^j &= \zadia^\frac{1}{2} \bar{C}^j_R,\\
\alpha &= Z_\alpha \alpha_R,
\end{align*}
\end{minipage}
\begin{minipage}{.19\textwidth}
\begin{align*}
V_\mu^\gamma &= \tilde{Z}_V^{\frac{1}{2}} V_{\mu R}^\gamma,\\
C^\gamma &= \tilde{Z}_{C}^\frac{1}{2} C^\gamma_R,\\
\lambda &= Z_\lambda \lambda_R,
\end{align*}
\end{minipage}
\begin{minipage}{.19\textwidth}
\begin{align*}
\omega^a &= \zogh^\frac{1}{2} \omega^a_R,\\
\bar{C}^\gamma &= \tilde{Z}_{\bar{C}}^\frac{1}{2} \bar{C}^\gamma_R,\\
\xi &= Z_\xi \xi_R,
\end{align*}
\end{minipage}
\begin{minipage}{.19\textwidth}
\begin{align*}
\bar{\omega}^a &= \zogh^\frac{1}{2} \bar{\omega}^a_R,\\
g&= Z_g g_R,\\
\phantom{x}&\phantom{=}
\end{align*}
\end{minipage}
all divergences can be absorbed. We define the renormalization group functions for any fields $\Phi$, parameters $\zeta$ and the gauge coupling $g^2$ according to,

\begin{minipage}{.32\textwidth}
\begin{align*}
\gamma_\Phi &=\frac{1}{2} \mu \frac{\partial }{\partial \mu}\log Z_\Phi,
\end{align*}
\end{minipage}
\begin{minipage}{.32\textwidth}
\begin{align*}
\gamma_\zeta &=-  \mu \frac{\partial }{\partial \mu}\log Z_\zeta,
\end{align*}
\end{minipage}
\begin{minipage}{.32\textwidth}
\begin{align*}
\beta_{g^2} &=  -2 g_R^2 \ \mu \frac{\partial}{\partial \mu} \log Z_g ,
\end{align*}
\end{minipage}
where $Z^{(1)}$ denote the respective one-loop parts. Dropping the subscript $R$ for the renormalized quantities hereafter, we obtain the following set of renormalization group functions,

\begin{minipage}{.5\textwidth}
\begin{align*}
\gamma_X &=- \frac{g^2}{(4\pi)^2}\frac{N}{2}\left( \frac{17}{6}-\frac{\xi}{2}-\frac{\alpha+(N-2)\lambda}{N-1} \right),\\
\gamma_V &= -\frac{g^2}{(4\pi)^2}\left(\frac{13N+9}{6}-\frac{\lambda}{2}(N-1)\right),\\
\tilde{\gamma}_V&= -\frac{g^2}{(4\pi)^2}N\frac{11}{3},\\
\gamma_{\bar{\omega}}&=-\frac{g^2}{(4\pi)^2}\frac{N}{2}\left( 3-\frac{\alpha +(N-2)\lambda}{N-1}\right),\\
\gamma_{\bar{C}}&= \frac{g^2}{(4\pi)^2}\frac{1}{2}[N\xi+3-\lambda(N-1)],\\
\tilde{\gamma}_{\bar{C}}&= \frac{g^2}{(4\pi)^2}\frac{N}{2}(3+\xi), \\
\beta_{g^2} &= - \frac{g^4}{(4\pi)^2}N\frac{22}{3}.
\end{align*}
\end{minipage}
\begin{minipage}{.5\textwidth}
\begin{align}
\gamma_\xi&= \frac{g^2}{(4\pi)^2}N\left(\frac{4}{3} -\xi-\frac{3}{\xi}\right),\nonumber \\
\gamma_\lambda &=  \frac{g^2}{(4\pi)^2}\left(\frac{13N+9}{3}-\lambda(N-1)\right), \nonumber\\
\gamma_\alpha &=  \frac{g^2}{(4\pi)^2}N\frac{22}{3},\nonumber\\
\gamma_\omega &=  - \frac{g^2}{(4\pi)^2}\frac{N}{2}\left( 3-\frac{\alpha +(N-2)\lambda}{N-1}\right),\nonumber \\
\gamma_C &=-\frac{g^2}{(4\pi)^2}\left[\frac{N}{2}(3+\xi)-\lambda(N-1)\right],\nonumber\\
\tilde{\gamma}_C&=-\frac{g^2}{(4\pi)^2}\frac{N}{2}(3+\xi),\nonumber\\
\phantom{x}&\phantom{x}\phantom{x} \label{1LoopRen}
\end{align}
\end{minipage}
Next, we analyse the composite operator as defined in \eqref{operator}. The (on-shell) BRST transformation reads,
\begin{equation}
\delta_B \mathcal{O} = \partial^\mu \left(X_\mu^a \omega^a \right),
\end{equation}
which is nothing but a total derivative and we thus obtained the BRST invariance. To render the operator physically meaningful, we must prove its multiplicative renormalizability. This is \textit{a priori} not clear, because all condensates with the same quantum number can mix upon renormalization. The entries of the related mass mixing matrix are determined by inserting the condensates in the various two-point functions and requiring cancellation of the appearing divergences. We obtained the following shape,
\begin{align}
&\mmo{\left[\frac{1}{2}X_{\mu\ R}^a X^\mu_{a\ R} \right] \\ \left[\frac{1}{2}V_{\mu\ R}^j V^\mu_{j\ R} \right] \\ \left[i \omega^a_R \bar{\omega}^a_R\right] \\ \left[i C^j_R \bar{C}^j_R \right] \\ \left[\frac{1}{2}V_{\mu\ R}^\gamma V^\mu_{\gamma \ R}\right] \\  \left[i C^\gamma_R \bar{C}^\gamma_R \right]} =\mmo{1-Z_{11}^{(1)} & 0 & -Z_{13}^{(1)} & 0 &0 & 0 \\- Z_{21}^{(1)} & 1-Z_{22}^{(1)} & -Z_{23}^{(1)} & 0 & 0& 0\\ -Z_{31}^{(1)} & 0 & 1-Z_{33}^{(1)} & 0 & 0 & 0 \\ 0 & -Z_{42}^{(1)} & 0 & 1 & 0 & 0 \\ -Z_{51}^{(1)} & 0 & -Z_{53}^{(1)} & 0 & 1 & 0 \\ 0 & 0 & 0 &0 & 0 & 1} \mmo{\left[\frac{1}{2}X_\mu^a X^\mu_a \right]_R \\ \left[\frac{1}{2}V_\mu^j V^\mu_j \right]_R \\ \left[i \omega^a \bar{\omega}^a\right]_R \\ \left[i C^j \bar{C}^j \right]_R \\ \left[\frac{1}{2}V_\mu^\gamma V^\mu_\gamma \right]_R \\ \left[i C^\gamma \bar{C}^\gamma \right]_R} .\label{invrenmatrix}
\end{align}
The requirement of multiplicative renormalizability $\mathcal{O}=Z^{1/2}_\mathcal{O}\mathcal{O}_R$ leads at one-loop level to the following condition,
%\begin{align}
%\frac{Z_X}{2} X_\mu^a X^{\mu}_a -i \xi_R Z_\xi \left(Z_\omega Z_{\bar{\omega}}%\right)^{1/2} \omega^a_R \bar{\omega}^a_R \overset{!}{=}Z_\mathcal{O}^{1/2} \left(\left[\frac{1}{2}X_\mu^a X^\mu_a \right]_R -\xi_R \left[i \omega^a \bar{\omega}^a\right]_R \right)
%\end{align} 
\muu{
-Z_{11}^{(1)}+Z_X^{(1)}+\xi Z_{31}^{(1)}\overset{!}{=}Z_\xi^{(1)}-Z_{33}^{(1)}+Z_\omega^{(1)}+\frac{1}{\xi}Z_{13}^{(1)}.\label{condi1}
}
Eq. \eqref{condi1} is found to be satisfied and yields the following anomalous dimension of the composite operator,
\muu{
\gamma_\mathcal{O}=\frac{\mu}{\mathcal{O}_R}\frac{\partial \mathcal{O}_R}{\partial \mu}=-\frac{1}{2}\mu \frac{\partial}{\partial \mu}\log Z_\mathcal{O}=\frac{g^2}{(4\pi)^2}\frac{N}{3}(13-3\xi).\label{compositeano}
}
As expected, all results agree for $N=2$ with the MAG \cite{Kondo:2000zva,Ellwanger:2002sj,Shinohara:2001cw,Dudal:2004rx}.

\section{Application of the LCO formalism}

The introduction of the composite operator and a corresponding source term $\int_x J \mathcal{O}$ leads to new divergences quadratic in the source, which must be taken care of. This is done within the Local Composite Operator (LCO) formalism \cite{Verschelde:1995jj,Knecht:2001cc}. We give a short overview of the main arguments in this formalism and state the results when applied to our theory. The LCO formalism is centred around the introduction of a new auxiliary parameter $\zeta$ together with a counterterm to remove the mentioned divergences quadratic in the source,

\muu{
\mathcal{L}_{\text{LCO}} = \frac{1}{2} (\zeta + \delta \zeta) J^2. \label{addpiece}
}
Assuming the RG invariance of $\mathcal{L}_{\text{LCO}}$, one can derive a differential equation to determine the auxiliary parameter under the assumption that it depends on the RG scale $\mu$ only independently through $g^2(\mu)$ and $\xi(\mu)$\footnote{We set $\alpha =\lambda =0$ hereafter as they are fixed points according to Eq. \eqref{1LoopRen}.},
\muu{
\left[2\epsilon + 2 \gamma_\mathcal{O}-\beta_{g^2}\frac{\partial}{\partial g^2}- \xi \gamma_\xi \frac{\partial}{\partial \xi} \right](\zeta +\delta \zeta)=0,\ \ \ \zeta(g^2, \xi)= \frac{\zeta_0}{g^2}+\hbar \zeta_1 + \hbar^2 \zeta_2 g^2 +\dots,
}
which can be solved order by order. Note, however, that $(n+1)$-loop results are necessary to determine $\zeta$ to $n$-loop level. The results from the previous section allow us to obtain the tree-level part $\zeta_0$ as,
\muu{
\zeta_0 =  \frac{2(N-1)}{N} \xi + C (4\xi - 3 \xi^2 -9),\label{kappa0sol}
}
which is enough to argue the existence of the condensate, see below. Next, a Hubbard-Stratonovich transformation is performed by inserting the identity,
\muu{
1= \int \mathcal{D}\sigma \ \ \text{exp}\left[-i\frac{1}{2g^2 \zeta}( \sigma - g \mathcal{O}-g\zeta J)^2\right].\label{hubbard}
}
which removes the $J^2$ term in the bare Lagrangian \eqref{addpiece} which otherwise would spoil the usual construction of the Schwinger functional. Moreover, the transformation introduces a tree-level mass term for the coset gluon and ghost, provided the auxiliary field $\sigma$ develops a non-zero vacuum expectation value. In fact, from the one-loop effective potential,
\muu{
V(\sigma) =& \frac{\sigma^2}{2\zeta_0}-\frac{\zeta_1}{2\zeta_0^2}g^2\sigma^2 - \frac{3}{64 \pi^2} 2(N-1) \frac{g^2 \sigma^2}{ \zeta_0^2}  \left( \frac{5}{6} - \log \left[ \frac{g \sigma}{\zeta_0 \bar{\mu}^2} \right] \right)\nonumber \\
&+\frac{1}{64 \pi^2}2(N-1)\frac{\xi^2 g^2 \sigma^2}{ \zeta_0^2}  \left( \frac{3}{2}-\log\left[ \frac{\xi g\sigma}{\zeta_0 \bar{\mu}^2} \right]\right); \ \ \bar{\mu}^2=4\pi\mu^2e^{-\gamma},
} 
we find a non-trivial extremum given by,
\muu{
m_X^2=\frac{g\sigma_*}{\zeta_0 }=\bar{\mu}^2\ \text{Exp}\left[-\frac{1}{(3-\xi^2)}\left(\frac{H_1}{g^2}+H_2 \right) \right], \label{minimum}
}
with
\muu{
H_1(\xi,\zeta_0)=\frac{32\pi^2}{2(N-1)}\zeta_0;\ \ \ \ 
H_2(\xi,\zeta_1)=-\left( \frac{32\pi^2}{2(N-1)}\zeta_1 +1+\frac{1}{2}\xi^2 \log \xi^2 -\xi^2  \right).\nonumber\\
\phantom{x}
}
The tree-level part of the potential must remain positive in order for the theory to be well-defined. Thus, we require $\zeta_0(\xi) >0$, at least for all $\xi$ in the physical region around $\xi=0$, see the discussion in the final section. This fact finally motivates a choice for the integration constant $C$ in Eq. \eqref{kappa0sol}. For convenience, we fix it to $C_0=-\frac{1}{11}\frac{N-1}{N}$. The restriction of $\xi$ to small values can also be deduced from the requirement of asymptotic freedom. In fact, requiring $m_X^2 \to 0$ for $g^2 \to 0$ implies that $H_1<0$ and thus using $\zeta_0 >0$ we infer $\xi^2 < 3$. The results obtained are all in agreement with the existing literature \cite{Dudal:2004rx}. Finally, enforcing $\xi =0$ from our interpretation of the reduction condition, the mass becomes proportional to the QCD scale,
\muu{
m_X^2=e^{H_2(\xi=0,\zeta_1)}\Lambda_{QCD}^2; \ \ \  \Lambda_{QCD}=\bar{\mu}\ \text{Exp} \left[-\int^g \frac{dg'}{\beta_g(g')} \right],
}
and thus, assuming that $\xi$ only varies slowly with $\mu$ around $\xi=0$, RG invariant. The fact that the vacuum energy corresponding to Eq. \eqref{minimum},
\muu{
V(\sigma_*)&=-(3-\xi^2)\frac{2(N-1)}{128\pi^2}\frac{g^2\sigma_*^2}{\zeta_0^2}=-(3-\xi^2)\frac{2(N-1)}{128\pi^2}m_X^4,
}
is negative under the derived condition $\xi^2<3$ renders the condensate to be energetically favoured, 

\section{Summary and conclusion}

In this talk we examined a novel decomposition of Yang-Mills theory with respect to the $H^{min}=U(N-1)$ stability group, in contrast to the well-known case of $H^{max}=U(1)^{N-1}$, which is related to the MAG. From a thorough one-loop analysis of this theory we obtained the full set of renormalization group functions. It turned out that in order to render the theory renormalizable, the stability group has to be further decomposed, $H^{min}=SU(N-1)\times U(1)$, where the $U(1)$ part corresponds to the center of $H^{min}$. Consequently, these degrees of freedom behave in the same way as the residual degrees of freedom in the MAG, in contrast to the $SU(N-1)$ sector. In fact, for $N=2$ our new option is identical with the MAG and in this limit our results are consistent with the exisiting literature. In particular, we recover the unpleasant feature of the ``gauge fixing'' parameter $\xi$ related to the fixing of the coset degrees of freedom having no fixed point.

In a next step, we proceeded to introduce a BRST invariant dimension two composite operator $\mathcal{O}=\Tr_{G/H^{min}} \left( \mathcal{X}_\mu \mathcal{X}^\mu - 2 i \xi \mathcal{C}\bar{\mathcal{C}}\right)$ in order to investigate the mass generation for certain gluon degrees of freedom from the viewpoint of the infrared Abelian dominance, a key concept within the dual superconductivity picture for quark confinement. The counterpart of this operator has already been studied within the MAG and the goal was to clarify whether this construction works in the novel option as well. After proving the BRST invariance and the multiplicative renormalizability of this operator at one-loop level, we applied the LCO formalism to discuss the existence of the condensate. The benefit of this formalism is twofold. First, if a source term for the composite operator is introduced, new divergences quadratic in the source are automatically generated. These divergences are removed within this formalism. Second, the accompanying bare term quadratic in the source which spoils the usual construction for the generating functional is removed by a Hubbard-Stratonovich transformation. The related auxiliary field $\sigma$ generates the desired mass term for the coset degrees of freedom once it develops a non-zero vacuum expectation value. In fact, the study of the one-loop effective potential $V(\sigma)$ revealed the existence of a non-trivial extremum $
\sigma^*$ which, however, explicitly depends on $\xi$ as does the corresponding vacuum energy $V(\sigma^*)$. On the other hand, the decomposition under consideration actually is just a consequence of approximating the more general situation of the reformulated Yang-Mills theory in the new variables by fixing the color-field. Within this reformulation, the reduction condition must be imposed rigorously. To say it in the language of the Fadeev-Popov procedure we are therefore inclined to set $\xi=0$ by hand in order to incorporate the reduction condition as a Delta function. Another fact that supports this picture is derived from the requirement of asymptotic freedom. If the induced coset gluon mass is required to vanish in the limit $g^2 \to0$, the restriction $\xi^2<3$ is obtained, i.e., $\xi$ must be small also from this perspective. Actually, just by using this inequality we found that $V(\sigma^*)$ is negative and the condensate is energetically favoured. Therefore, we conclude that the condensate exists also in our new decomposition.

\section*{Acknowledgements}

We thank J.A. Gracey and D. Dudal for useful discussions. M. W. was supported by the Chiba University SEEDS Fund (Chiba University Open
Recruitment for International Exchange Program) and by the Ministry of Education,
Culture, Sports, Science and Technology, Japan (MEXT
scholarship). R. M. was supported by Grant-in-Aid for Japan Society for the Promotion of Science (JSPS)
Research  Fellow  Grant  No.  17J04780.  S. N.  thanks
Nakamura Sekizen-kai for a scholarship. K.-I. K. was
supported by Grant-in-Aid for Scientific Research, JSPS
KAKENHI Grant No. 15K05042.

\end{document}